# CP VIOLATION IN TOP PHYSICS AT THE NLC

D. ATWOOD

*Theory Group, National Jefferson Lab,*

A. SONI[a]

*Theory Group, Brookhaven National Laboratory, Upton, NY 11973*

Top quark is extremely sensitive to non-standard CP violating phases. General strategies for exposing different types of phases at the NLC are outlined. SUSY phase(s) cause PRA in $t \to Wb$. The transverse polarization of the $\tau$ in the reaction $t \to b\tau\nu$ is extremely sensitive to a phase from the charged Higgs sector. Phase(s) from the neutral Higgs sector cause appreciable dipole moment effects and lead to sizable asymmetries in $e^+e^- \to t\bar{t}H^0$ and $e^+e^- \to t\bar{t}\nu_e\bar{\nu}_e$.

## 1 Introduction

The standard model (SM) with three families and the CKM phase gives a natural explanation for the size of the observed CP asymmetry in the neutral kaon complex. It also strongly suggests the presence of large CP asymmetries in $B$-decays making $B$-physics very suitable for precision extraction of the CKM phase and a quantitative check of the unitarity triangle. Furthermore, the minimum SM with $m_t \sim 175$ GeV $\gg m_b, m_d, m_s$ also leads to the conclusion that CP violation effects in top production and decays have to be extremely small. However, extensions of the SM, invariably lead to new CP violating phase(s). Indeed in most extensions new CP violating phases appear rather naturally. In fact getting rid of non-standard phases in extensions of the SM may be as unnatural as it is to get rid of the CKM phase from the three generation SM. Since it is widely believed that the SM cannot account for baryogenesis CP violation from new physics is very likely a necessity. Therefore, it is important for us to seek optimal strategies to expose different types of non-standard CP violating phases. Based on studies that are so far available the prominent effects of different types of phases appears to be the following:

1. Dipole moment of the top quark could be large enough, most effectively due to phase(s) from the neutral Higgs sector, to be experimentally accessible.[1-4]

2. SUSY phase(s) cause interesting partial rate asymmetry (PRA) effects in decays of the type $t \to Wb$.[5-6]

3. The transverse polarization of the $\tau$ in the decay $t \to b\tau\nu$ is very sensitive to phase(s) originating from the charged Higgs sector.[7]

4. CP violating phase(s) due to $H^0$ exchanges cause large asymmetries[8,9] in $e^+e^- \to t\bar{t}H^0$ and in $e^+e^- \to t\bar{t}\nu_e\bar{\nu}_e$.[10,11]

## 2 Top Polarimetry

The fact that the top is so heavy has the important consequence that the top quark does not bind into hadrons. All CP violation in the top quark is therefore direct CP. Spin of the top quark becomes a very important observable and its decays become very effective analyzers of its spin.[12] The polarization of the top-quark is strongly correlated to the directions of the momentum of the charged lepton in $t \to b\ell\nu_\ell$, or to the $W$-momentum in $t \to bW$, or to the direction of the most energetic jet in $t \to b + 2$jets.[13,14] The ability to track the top spin thus plays a crucial role in CP violation tests involving the top quark.

## 3 Dipole Moment of the Top-Quark

In many extensions of the SM the top quark can acquire dipole moment at one loop order.[1,2] For example, CP violation phase from a neutral Higgs sector can cause top dipole moment of order $10^{-20} e-$cm i.e. about ten orders of magnitude more than in the SM wherein one needs to go to at least two loop order and it is expected to be $\lesssim 10^{-30} e$-cm. Much attention has recently been given to detection of the dipole moment form factor.[1,3-4] It should be clear that the form factor has not only a real part but there is also an imaginary part arising at $q^2 \geq 4m_t^2$. Since the life-time of the top quark is so short form factor extraction requires simultaneous studies of production and subsequent decay. A multitude of (CP Violating) observables that are $T_N$-odd or $T_N$-even can be constructed using beam polarization, beam momenta, $t, \bar{t}$ momenta and/or momenta of various decay products.[1,3-4] Under these circumstances it is natural to search for an optimal observable,[3] namely, one that is the most efficient for the determination of (e.g.) the real or the imaginary part of the dipole moment form factor for a fixed value of $q^2$. Separating the differential cross-section into a CP conserving ($\Sigma_0$) and a CP violating ($\Sigma_1$) piece[4]

$$\sigma(\phi) = \Sigma_0(\phi) + \lambda \Sigma_1(\phi) \quad (1)$$

then the optimal observable, for determination of the dipole moment $\lambda$ is given by[4]

$$0_{opt} = \Sigma_1(\phi)/\Sigma_0(\phi) \quad (2)$$

[a]Presenter

Studies have shown that the anticipated luminosity at the NLC[15-17] of $5 \times 10^{33}$ cm$^{-2}$s$^{-1}$ [14] should be able to give limits on $\lambda$ to the level of $\sim 10^{-19}$e-cm at one sigma.[4] Thus observation of the top dipole moments that are expected in theoretical extensions would present somewhat of a challenge but their feasibility cannot be excluded.

## 4 PRA and the Supersymmetric Phase[5,6]

Two studies indicate that, in decays of the type $t \to Wb$, SUSY phase(s) can lead to PRA up to a few %. Experimentally there are two ways to look for these effects. Firstly PRA can be searched for by comparing $V_{tb}$ with $V_{\bar{t}\bar{b}}$. The expectations are that at the NLC, $V_{tb}$ can be measured to a precision of a few %.[15-17] So a difference between $V_{tb}$ and $V_{\bar{t}\bar{b}}$ less than about 10% would be rather difficult to detect.

Incidentally, a second way to search for a PRA would be to determine the $BR$ for decays that are not accompanied by a beauty quark i.e. $BR$ into "ugly decays" of the top quark. This would yield the effective CKM angle $V_{tx}$ into non-beauty decays. Separation into decays that contain a prompt $d$-quark versus a prompt $s$-quark is probably quite unrealistic. However for the purpose of PRA this separation is not necessary. Since PRA in $t \to Wb$ can only arise if there are (non-standard) decays of the top, which provide the needed absorptive parts, it means that CKM unitarity must be violated by $V_{tx}$ i.e.

$$|V_{tx}|^2 \neq [1 - |V_{tb}|^2] \quad (3)$$

For this type of precision studies at the NLC experiments must be able to search for such "ugly decays" (i.e. top decays not accompanied by prompt beauty) with high efficiency. Of course efficiency in $b$-tagging is also extremely important.

## 5 Transverse Polarization of the $\tau$ due to a Charged Higgs Phase.[7]

The transverse polarization of the $\tau$ in the decay $t \to b\tau\nu$ is extremely sensitive to the presence of a CP violating phase from the charged Higgs sector. The effects are large as they originate from tree level interferences and also the $W$-propogator is on-shell and causes a "resonance" enhancement. There is a CP-odd, $T_N$-odd transverse polarization asymmetry sensitive to the real part of the $W$-propogator and a CP-odd, $T_N$-even transverse polarization asymmetry that is driven by the imaginary part of the resonant $W$-propogator. The second type of asymmetry does not occur in $K \to \pi\mu\nu$ or $B \to D(D^*)\tau\nu\ldots$. In top decays both types of asymmetries are quite large. For $m_H \sim 200$ GeV they can be as big as 50% and for $m_H \sim 400$ GeV they can be in the range 5–20%. At the NLC, at $\sqrt{s} = 500$ GeV, with a luminosity of $5 \times 10^{33}$ cm$^{-2}$s$^{-1}$ we may be able to probe asymmetries larger than about 6% with three sigma sensitivity. Clearly, the ability to measure the polarization of the $\tau$ would be a very important feature in experiments at the NLC.

## 6 Neutral Higgs CP in $e^+e^- \to t\bar{t}H^0$.

CP violation phase due to a neutral Higgs sector leads to very interesting effects that occur through the interference of two tree graphs (see Fig.1).[8] Without using beam polarization and/or $t, \bar{t}$ spin, there is only one triple correlation that is possible, i.e. $(\vec{p}_{e^+} - \vec{p}_{e^-}) \cdot (\vec{p}_t \times \vec{p}_{\bar{t}})$. The resulting, $T_N$-odd, asymmetries are in the 10–25% range. The cross section is about a few $fb$. Thus these asymmetries should be observable if the projected luminosity of $2 \times 10^{34}$ cm$^{-2}$s$^{-1}$ could be reached for a 1 TeV machine. The method is most suitable for a relatively light Higgs, $m_{H^0} \lesssim 2m_W$, as then the $BR(H_0 \to b\bar{b})$ is substantial. Gunion et al.[9] have extended the study of this reaction. With the use of generalized optimal observables they claim that at $E = 1.5$ TeV the NLC would enable determination of the basic couplings for $H\bar{t}t$, $H\bar{t}\gamma_5 t$ and $HZZ$ vertices and thereby clarify the nature of the Higgs boson.[9]

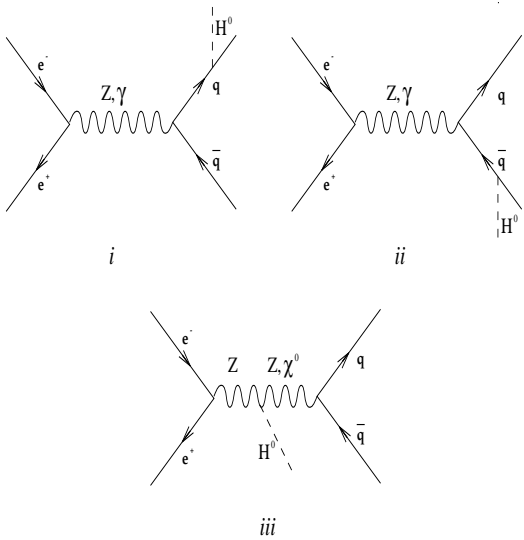

Figure 1: Tree-level Feynman diagrams contributing to CP violation in $e^+e^- \to t\bar{t}H^0$ in the two Higgs doublet model

## 7 Neutral Higgs Phase in $e^+e^- \to \nu\bar{\nu}W^+W^-$.

This reaction is extremely interesting as it also leads to a $T_N$-odd polarization asymmetry that arises at tree-level.[10] Furthermore, the resonant Higgs propagator leads to enhanced effects resulting in the possibility of substantial asymmetries $\sim$ 10–50%. The relevant Feynman graphs are shown in Fig. 2.

Actually there are three types of polarization asymmetries that are of interest. To define these let us introduce, in the rest frame of the $t$, the basis vectors: $-\hat{e}_z \alpha(\vec{p}_{W^+} + \vec{p}_{W^-})$; $\hat{e}_y \alpha(\vec{p}_{W^+} \times \vec{p}_{W^-})$ and $\hat{e}_x = \hat{e}_y \times \hat{e}_z$. Let $P_j$ (for $j = x, y, z$) be the polarization of the $t$ along $\hat{e}_x$, $\hat{e}_y$ or $\hat{e}_z$. Similarly for $\bar{t}$. Combining the information from $t, \bar{t}$ we define the CP violating asymmetries:

$$A_x = \frac{1}{2}(p_x + \bar{p}_x); \qquad A_y = \frac{1}{2}(p_y - \bar{p}_y)$$
$$A_z = \frac{1}{2}(p_z + \bar{p}_z) \qquad (4)$$

So $A_x$ and $A_z$ are $T_N$-even requiring absorptive parts whereas $A_y$ is $T_N$-odd requiring real Feynman amplitude. Therefore $A_y$ arises from interference among the tree level graphs in Fig. 2. Near resonance $A_x$ also arises primarily from these tree graphs; the Higgs width provides the necessary absorptive parts. $A_z$, though, receives additional contributions from loop graphs.

In addition to CP violating polarization asymmetries there is also an interesting CP-even, $T_N$-odd asymmetry that is quite sizable. We should be able to use it to determine the Higgs width in SM as well as in its extensions.

The works of Ref. 10,11, on the reaction $e^+e^- \to t\bar{t}\nu_e\bar{\nu}_e$, are complementary. Ref. 10 deals with interference of a neutral Higgs exchange with the other SM graphs of Fig. 2. Ref. 11 ignores that contribution to CP violation and focuses primarily on interference between two neutral Higgs occurring in a 2HDM.

## 8 Summary

SM causes negligible CP violating effects in top physics whereas many types of extensions lead to large effects therein. The top can receive a large dipole moment from neutral Higgs CP. The latter can also cause significant asymmetries in $e^+e^- \to t\bar{t}H^0$ and $e^+e^- \to t\bar{t}\nu_e\bar{\nu}_e$ arising at tree level. SUSY phase(s) can cause PRA in $t \to Wb$. Phase(s) from the charged Higgs sector result in large transverse polarization of $\tau$ in $t \to b\tau\nu$.

There is thus the exciting possibility that the parameters needed for understanding of baryogenesis could be extracted or verified through CP

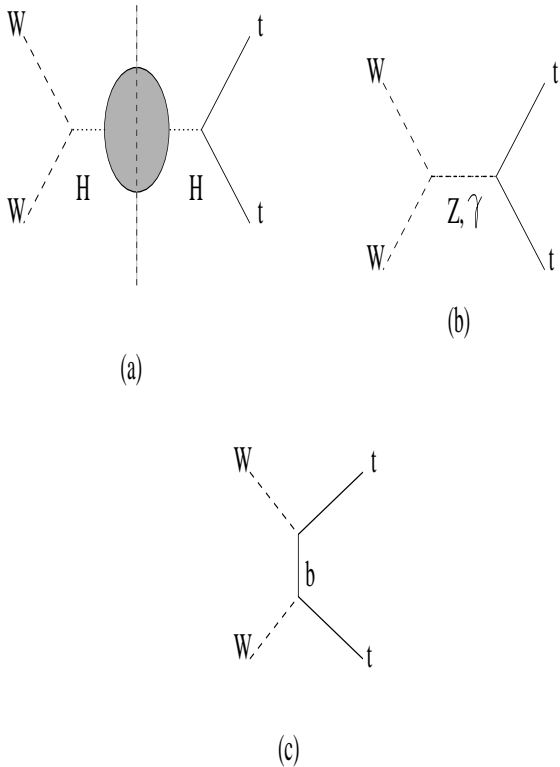

Figure 2: The Feynman diagrams that participate in the sub-process $W^+W^- \to t\bar{t}$. The blob in Fig.2a represents the width of the Higgs resonance and the cut across the blob is to indicate the imaginary part.

violation studies of the top quark in accelerator experiments.

### Acknowledgments

This research was supported in part by the US-DOE contracts DC-AC05-84ER40150 (CEBAF) and DE-AC-76CH00016 (BNL).

### References

1. W. Bernreuther, T. Schroöder and T.N. Pham, *Phys. Lett.* B **279**, 389 (1992).
2. A. Soni and R. Xu, *Phys. Rev. Lett.* **69**, 33 (1992).
3. C.R. Schmidt and M. Peskin, *Phys. Rev. Lett.* **69**, 410 (1992); W. Bernreuther and P. Overman, preprint hep-ph/9511256; F. Cuypers and S.D. Rindani, *Phys. Lett.* B **343**, 333 (1995)
4. D. Atwood and A. Soni, *Phys. Rev.* D **45**, 2405 (1992).
5. B. Grzadkowski and W.Y. Keung, *Phys. Lett.* B **319**, 526 (1993).
6. E. Christova and M. Fabbrichesi, *Phys. Lett.* B **320**, 229 (1994).
7. D. Atwood, G. Eilam and A. Soni, *Phys. Rev. Lett.* **71**, 492 (1993).
8. S. Bar-Shalom *et al.*, *Phys. Rev.* D **53**, 1162 (1996).
9. J.F. Gunion, B. Grzadkowski and X.-G. He, preprint UCD-96-14.
10. D. Atwood and A. Soni, preprint JLAB-TH-96-14.
11. A. Pilaftsis and M. Nowakowski, *Int. Jour. of Mod. Phy.* A **9**, 1097 (1994).
12. See e.g. C.R. Schmidt and M. Peskin, Ref. 3; G.L. Kane, G.A. Ladinsky and C.P. Yuan, PRD **45**, 124 (1992); D. Atwood, A. Aeppli and A. Soni, *Phys. Rev. Lett.* **69**, 2754 (1992).
13. B. Grzadkowski and J.F. Gunion, *Phys. Lett.* B **350**, 218 (1995).
14. D. Atwood and A. Soni, *Phys. Rev.* D **52**, 6271 (1995).
15. Proceedings of the Workshop on Physics and Experiments with Linear $e^+e^-$ Colliders, Eds. F. Harris, S. Olsen, S. Pakvasa and X. Tata, World Scientific, Singapore 1996.
16. Proceedings of the Workshop on Physics and Experiments with Linear Colliders, Eds. A. Miyamoto and Y. Fuji, World Scientific, Singapore, 1996.
17. H. Murayama and M. Peskin, preprint SLAC-PUB-7149 (hep-ex/9606003).